\documentclass[a4paper,11pt]{article}
\pdfoutput=1 
\usepackage{ifpdf}

\usepackage{jcappub} 

\usepackage[T1]{fontenc} 
\usepackage{color}
\usepackage{textcomp}

\usepackage{soul}
\usepackage{ulem}

\usepackage{import}
\usepackage{xcolor}
\usepackage{comment}

\title{Non-linear effects on the Cosmological Gravitational Wave Background anisotropies}
\author[a,b]{Alina Mierna,}
\author[a,b,c,d]{Sabino Matarrese,}
\author[a,b,c,d]{Nicola Bartolo,}
\author[e,f]{Angelo Ricciardone}

\affiliation[a]{Dipartimento di Fisica e Astronomia ``G. Galilei'',
Universit\`a degli Studi di Padova, via Marzolo 8, I-35131 Padova, Italy}
\affiliation[b]{INFN, Sezione di Padova,
via Marzolo 8, I-35131 Padova, Italy}
\affiliation[c]{INAF- Osservatorio Astronomico di Padova, \\ Vicolo dell’Osservatorio 5, I-35122 Padova, Italy}
\affiliation[d]{Gran Sasso Science Institute, Viale F. Crispi 7, I-67100 L’Aquila, Italy}
\affiliation[e]{Dipartimento di Fisica “Enrico Fermi”, Universit\`a di Pisa, Pisa I-56127, Italy}
\affiliation[f]{INFN sezione di Pisa, Pisa I-56127, Italy}

\begin{document}

\abstract{The Cosmological Gravitational Wave Background (CGWB) anisotropies contain valuable information about the physics of the early universe. Given that General Relativity is intrinsically nonlinear, it is important to look beyond first-order contributions in cosmological perturbations. In this work, we present a non-perturbative approach for the computation of CGWB anisotropies at large scales, providing the extension of the initial conditions and the Sachs-Wolfe effect for the CGWB, which encodes the full non-linearity of the scalar metric perturbations. We also derive the non-perturbative expression for three-point correlation of the gravitational wave energy density perturbation in the case of an inflationary CGWB with a scale-invariant power spectrum and negligible primordial non-Gaussianity. We show that, under such conditions, the gravitational wave energy density perturbations are lognormally distributed, leading to an interesting effect
such as intermittency. }

\maketitle

\section{Introduction}

The Cosmological Gravitational Wave Background (CGWB) provides a unique opportunity to observe the early universe, since the universe is transparent to gravitational waves (GWs) below the Planck energy scale. Recently, the PTA collaboration have claimed evidence for the detection of a Gravitational Wave Background (GWB) in the nHz frequency range~\cite{NANOGrav:2023gor, EPTA:2023fyk, Reardon:2023gzh}. Various possible interpretations of the signal of both astrophysical and cosmological origin have been proposed~\cite{NANOGrav:2023hvm, NANOGrav:2023hfp, EPTA:2023xxk, Franciolini:2023pbf, Franciolini:2023wjm, Ellis:2023tsl, Vagnozzi:2023lwo, Figueroa:2023zhu}, however the exact origin of the background is still unclear. Therefore, it is essential to have as many observables as possible to characterize the GWB in order to discriminate between different potential sources of GWs. Considering that the angular resolution of planned GW experiments will significantly improve, anisotropies of the GW energy density could provide a crucial information about the origin of the CGWB~\cite{Cornish:2001hg, Alba:2015cms,Contaldi:2016koz, LISACosmologyWorkingGroup:2022kbp, Bartolo:2019oiq, ValbusaDallArmi:2020ifo,Schulze:2023ich}.  

The Boltzmann equation describes the evolution of GWs in the limit of geometric optics, and, analogously to the Cosmic Microwave Background (CMB), can be used to obtain the GW energy density perturbation at the linear level~\cite{Contaldi:2016koz, Bartolo:2019oiq, Bartolo:2019zvb, Ricciardone:2021kel,LISACosmologyWorkingGroup:2022kbp, Bartolo:2019yeu,ValbusaDallArmi:2020ifo,Perna:2023dgg,Schulze:2023ich}. However, there is an alternative way to compute anisotropy of the CGWB on large scales. According to Liouville’s Theorem, the phase-space distribution function is constant along the trajectories of decoupled system. Hence, the perturbations in the graviton distribution function today are directly connected to the inhomogeneities imprinted at the moment of  GW production through their propagation. Differently from the CMB case, where the Boltzmann equation is necessary to account for the interactions of photons with other particle species before recombination, gravitons decouple at the Planck epoch and hence initial conditions are set at the very early time and further they propagate freely till the observation point. This approach was used to compute the second-order CMB anisotropies in~\cite{Pyne:1995bs, Mollerach:1997up}. Given that CMB photons are described by the Planck distribution, this results in the observed temperature anisotropies being related to the ones at last scattering through the ratio of the emitted and observed frequencies. 

The non-thermal nature of GWs makes it difficult to obtain the corresponding relation for the anisotropies in the GW energy density. Typically, the graviton distribution function is expanded as a leading isotropic component that is only a function of the comoving momentum plus a first-order anisotropic contribution. Given that General Relativity is a non-linear theory, it could be important to consider beyond first-order contributions to the anisotropy of the CGWB. Assuming the graviton distribution function can be parameterized by a general power-law form, we can connect the initial anisotropy of the CGWB to the one observed today going beyond perturbation theory. The CGWB anisotropies are generated both by the GW production mechanism and by the GW propagation through the perturbed universe. The initial anisotropy depends on the specific mechanisms that source GWs. We focus on the GWB generated by the quantum vacuum fluctuations of the metric during inflation. In this case, it is possible to obtain a non-perturbative generalization of initial conditions for the CGWB on large scales by recasting the large-scale scalar and small-scale tensor perturbations in such a way as to obtain a new metric, which describes, on large scales, a homogeneous and isotropic universe. Such a rescaling was employed for the CMB~\cite{Bartolo:2011wb, Boubekeur:2009uk} to account for second-order anisotropies and to compute the non-linear generalizations of the Sachs-Wolfe (SW) eﬀect~\cite{Bartolo:2005fp}. 
A similar formalism was also extended to include the integrated Sachs-Wolfe effect (ISW) in~\cite{Roldan:2017wvm}. We consider a flat spacetime, however a non-zero spatial curvature could enhance the non-linear effects on the CGWB anisotropies~\cite{Cai:2024dya}.

The non-linearity of the CGWB anisotropies would also affect higher-order correlation functions. Using the path-integral approach, we compute the 3-point correlation function (and corresponding bispectrum) of the GW energy density, which in the case of GWB generated by quantum tensor fluctuations during inflation with a scale-invariant power spectrum and negligible primordial non-Gaussianity will be exact and hold at any order in perturbation theory. 

The paper is organized as follows. In Section 2, we present the non-linear extension of the initial conditions and the SW effect for the CGWB. In Section 3, we express the CGWB anisotropy in terms of the comoving curvature perturbation and subsequently compute the bispectrum of the GW energy density. Section 4 contains our conclusions.

\section{The non-perturbative Sachs-Wolfe effect for the CGWB} 
In the geometric optics limit, gravitons can be regarded as massless, collisionless particles propagating along null geodesics of the perturbed background metric. We can define a phase-space distribution function for gravitons $f_{\rm GW}\left(\eta, \Vec{x}, \Vec{q}\right)$ as a function of the conformal time $\eta$, the position $\Vec{x}$ and the comoving momentum $\Vec{q}$ $(q= |\Vec{p}| a)$. The evolution of the distribution function for a given particle species is described by the Boltzmann equation. In the absence of collisions, the Boltzmann equation for the graviton distribution function takes the form
\begin{equation}
\frac{d f_{\rm GW}}{d \eta} = 0\,.
\end{equation}
Consequently, the graviton distribution function is constant along the trajectories of the system. In other words, the number of particles in a given element of phase space does not change with time. Since the phase-space distribution function is constant along the trajectories of the system, we can connect the graviton distribution function at the emission time $\eta_{\rm in}$ to the one computed at the present time 
$\eta_{0}$
\begin{equation}
\label{f_GW = const}
    f_{\rm GW}\left(\eta_{\rm in}, \Vec{x}_{\rm in}, \Vec{q}_{\rm in}\right) = f_{\rm GW}\left(\eta_{\rm 0}, \Vec{x}_{\rm 0}, \Vec{q}_{\rm 0}\right)  \, .
\end{equation}
One can directly relate the phase-space distribution function to the GW energy density at a given time and location by
\begin{equation}
    \rho_{\rm GW}\left(\eta, \Vec{x}\right)= \frac{1}{a^4(\eta)} \int d^3q \, q  \, f_{\rm GW}\left(\eta, \Vec{x}, \Vec{q}\right)  = \rho_{\rm crit} \int d\ln{q} \, \Omega_{\rm GW}\left(\eta, \Vec{x}, \Vec{q}\right)
\end{equation}
where $a(\eta)$ is the scale factor and $\Omega_{\rm GW}\left(\eta, \Vec{x}, \Vec{q}\right)$ is the spectral energy density. The average GW energy density is connected to the isotropic component of the graviton distribution function through  
\begin{equation}
\label{GW_av}
    \overline{\Omega}_{\rm GW} = \frac{4\pi}{\rho_{\rm crit}} \left(\frac{q}{a}\right)^4 \overline{f}(q)\, ,
\end{equation} 
and the anisotropic component can be taken into account by defining $\omega_{\rm{GW}}$ as
\begin{equation}
    \Omega_{\rm GW} = \int \, d^2\hat{n} \, \omega_{\rm GW}\left(\eta, \Vec{x}, \Vec{q}\right)/4\pi.
\end{equation}
We can then introduce the energy density perturbation of gravitons 
\begin{equation}
    \delta_{\rm{GW}} = \frac{\omega_{\rm{GW}} -  \bar{\Omega}_{\rm GW}\left(q\right)}{ \bar{\Omega}_{\rm GW}\left(q\right)} = \frac{f - \overline{f}}{\overline{f}} = \frac{\delta f}{\overline{f}} \,.
\end{equation}
The average GW energy density is typically parametrized by a power-law frequency profile~\cite{LISACosmologyWorkingGroup:2024hsc}. From eq. \eqref{GW_av}, the isotropic component of the graviton distribution function can be written as
\begin{equation}
\begin{split}
    \overline{f}_{\rm GW}\left(q\right) &\propto
    q^{n_{\rm gwb}-4}\,, 
\end{split}
\end{equation}
where $n_{\rm gwb}$ is the GWB spectral index, 
\begin{equation}
    n_{\rm gwb}(q) =  \frac{ \partial \ln \, \bar{\Omega}_{\rm{GW}}\left(\eta, q\right)}{\partial  \ln  q}\,.
 \end{equation}
Expanding the graviton distribution function to first order in perturbations, we get
\begin{equation} 
f_{\rm GW}(\eta, \Vec{x}, \Vec{q})= \overline{f}_{\rm GW}(q)\left(1 - \left[n_{\rm gwb}(q)-4\right]\Gamma(\eta, \Vec{x}, \Vec{q}) \right)\, ,
\end{equation}
where $\Gamma(\eta, \Vec{x}, \Vec{q})$ is the dimensionless perturbation of the energy of the gravitons at first order. We can generalize this expression if we assume that it is a first-order approximation of the total graviton distribution function that can be parametrized by a power law with the additional exponential prefactor, which denotes the anisotropic component 
\begin{equation}
\begin{split}
    f_{\rm GW}\left(q, \Gamma\right) &\propto\left[qe^{-\Gamma}\right]^{n_{\rm gwb}-4}\, , 
\end{split}
\end{equation}
where the directional dependence is encoded in the function  
$\Gamma(\eta, \Vec{x}, \Vec{q})$ at any order. Substituting this expression in eq. \eqref{f_GW = const}, we obtain the non-perturbative expression for the energy density perturbation of gravitons today 
\begin{equation}
\label{ratio}
    e^{\Gamma_{0}} =  \frac{q_{0}}{q_{\rm in}}e^{\Gamma_{\rm in}} \, .
\end{equation} 
The total GW energy density perturbation includes the intrinsic inhomogeneity of the GW production mechanism and the contribution from the propagation of gravitons through the large-scale perturbations of the metric, which is provided by the ratio of the comoving momenta at the departure and arrival points.

In what follows, we will consider the CGWB generated by the quantum fluctuations of the metric during inflation and derive the initial conditions for this scenario. Then, we will solve the time component of the geodesic equation to obtain the ratio of the comoving momenta.

\subsection{Initial conditions}
The line element of the perturbed Friedmann-Lemaitre-Robertson-Walker (FLRW) metric in the Poisson
gauge reads
\begin{equation}
ds^2 = a^2(\eta)\left[-e^{2\Psi}d\eta^2 +\left(e^{-2\Phi}\delta_{ij}+\gamma_{ij}\right)dx^idx^j\right],
\end{equation}
where $\Phi(\eta, \Vec{x})$ and $\Psi(\eta, \Vec{x})$ represent the scalar perturbations and $\gamma_{ij}$ are the transverse-traceless tensor fluctuations. Beyond linear order, scalar, vector and tensor perturbations are coupled,, and vector modes could be present. However, we can neglect vector perturbations, since linear vector perturbations are not generated in standard scenarios of inflation as well as in other early universe mechanisms that generate cosmological perturbations, while at higher orders they will contain gradients of scalar modes and thus they can be neglected on large scales.

In the shortwave approximation, it is possible to split tensor perturbations into the small-scale modes that we identify as GWs $h_{ij}$ and the large-scale tensor perturbations $H_{ij}$~\cite{ValbusaDallArmi:2023nqn,ValbusaDallArmi:2024hwm}. The wavelength of GWs is much smaller than the scales on which the background perturbations vary, and hence $\gamma_{ij} = h_{ij} + H_{ij}$.
We neglect the large-scale tensor perturbations at all orders, since
already at linear
order the primordial tensor spectrum is suppressed with respect to
the scalar one \cite{Galloni:2022mok}.

Assuming that the GWs are well inside the Hubble radius at the time when we set initial conditions, we can neglect the spatial derivatives of the large-scale scalar perturbations of the metric, $\Phi$ and $\Psi$, since at this time their wavelengths are outside the horizon. In this case, we can make a change of coordinates in such a way as to absorb the superhorizon scalar modes and we can work with a metric of a universe which resemble to be a homogeneous and isotropic universe plus GWs. We can split the scalar perturbations into 
\begin{equation}
    \begin{split}
       \Phi &= \Phi_{\ell} + \Phi_s \, ,\\
        \Psi &= \Psi_{\ell}+ \Psi_s\, .
    \end{split}
\end{equation}
where subhorizon modes contribute only to $\Phi_{s}$ and $\Psi_{s}$. Then, $\Phi_{\ell}$ and $\Psi_{\ell}$ contain only the superhorizon modes and therefore their spatial derivatives are negligible. We can assume that they are only functions of the conformal time $\eta$. For this reason, we can absorb these large-scale perturbations of the metric by the following redefinition
\begin{equation}
\label{redef}
    \begin{split}
        \Tilde{a}&= e^{-\Phi_{\ell}}a\\
        d\Tilde{\eta}&= e^{\Psi_{\ell}+ \Phi_{\ell}}d\eta\\
        \Tilde{h}_{ij}&=e^{2\Phi_{\ell}}h_{ij} \, .
    \end{split}
\end{equation}
The small-scale tensor modes also have to be redefined to avoid mixing of the scalar and tensor perturbations in the definition of GWs. From now on, we drop the subscript $\ell$ for simplicity and refer only to superhorizon modes. The new metric then describes a homogeneous and isotropic universe on large scales plus a modulated small-scale (i.e high-frequency) tensor mode~\footnote{{Equivalently we can say that the new metric contains also a large-scale tensor mode that arise from a large-scale scalar mode modulating a small-scale (high-frequency) tensor mode, that in fact cannot be reabsorbed into a FLRW metric.}}
\begin{equation}
ds^2 = \Tilde{a}^2(\Tilde{\eta})\left[-d\Tilde{\eta}^2 +(\delta_{ij}+\Tilde{h}_{ij})dx^idx^j\right]\,.
\label{eq:red_metric}
\end{equation}
Thus, on superhorizon scales the universe can be considered as a collection of regions of the size of the Hubble radius that evolve independently as unperturbed patches with the new metric~\cite{PhysRevD.42.3936}.

Now we can compute the energy density of GWs in the new metric. The energy-momentum tensor of GWs defined in terms of covariant derivatives of the GWs reads~\cite{Isaacson:1967sln}
\begin{equation}
    T_{\mu\nu}^{\rm GW} = \frac{1}{32\pi G}\left\langle\mathcal{D}_\mu \Tilde{\gamma}_{\alpha\beta}^{\rm GW}\mathcal{D}_\nu\Tilde{\gamma}^{\rm GW\,\alpha\beta}\right\rangle \, .
    \label{def:T_Isaacson}
\end{equation}
The GWs in this prescription are identified as
\begin{equation}
    \begin{split}        \Tilde{\gamma}_{ij}^{\rm GW} \equiv & \Tilde{a}^2 \Tilde{h}_{ij} \, , \\
        \Tilde{\gamma}^{{\rm GW}\, ij} \equiv &  \Tilde{g}^{ik}\Tilde{g}^{jl}\Tilde{\gamma}_{kl}^{\rm GW} = \frac{1}{\Tilde{a}^2}\Tilde{h}^{ij}\, .
    \end{split}
\end{equation}
Then the energy density of GWs is given by
\begin{equation}
\label{GWdensity}
    \begin{split}
        \rho_{\rm GW}  =& \frac{1}{32\pi G \Tilde{a}^2  }\left\langle \Tilde{h}_{ij}^\prime\Tilde{h}^{ij\,\prime}\right\rangle   \, .
    \end{split}
\end{equation} 
Since the large-scale scalar perturbations have been reabsorbed into the metric, this expression corresponds to the total energy density of GWs. We can rewrite it as
\begin{equation}
    \begin{split}
        \rho_{\rm GW}  =& \frac{1}{32\pi G \Tilde{a}^2  }\left\langle \frac{\partial \Tilde{h}_{ij} }{\partial\Tilde{\eta}}\frac{\partial \Tilde{h}^{ij\,}}{\partial\Tilde{\eta}}\right\rangle = \frac{1}{32\pi G a^2  e^{-2\Phi} }\frac{1}{e^{2\Psi + 2\Phi} } e^{4\Phi}\left\langle \frac{\partial h_{ij} }{\partial \eta}\frac{\partial h^{ij\,}}{\partial \eta}\right\rangle  \\
        &= \frac{1}{32\pi G a^2} e^{-2\Psi + 4\Phi}\left\langle h_{ij}^\prime h^{ij\, \prime}\right\rangle  \, .
    \end{split}
\end{equation} 
This expression holds at any order in perturbation theory. To make a connection with the the standard perturbative approach, we can expand the exponential prefactor, which at first order results in
\begin{equation}
    \begin{split}
        \rho_{\rm GW}  
        &= \frac{1-2\Psi + 4\Phi}{32\pi G a^2} \langle h_{ij}^\prime h^{ij\,\prime }\rangle  \, .
    \end{split}
\end{equation}
which corresponds to the initial conditions for the CGWB generated by the quantum fluctuations of the metric obtained in~\cite{ValbusaDallArmi:2024hwm}. At this point, it is immediate to connect the GW energy density perturbation to the perturbation of the graviton distribution function 
\begin{equation}
   \delta_{\rm GW, \, \rm in }= \frac{e^{-2\Psi + 4\Phi}}{\langle e^{-2\Psi + 4\Phi} \rangle} -1 = \frac{e^{-\Gamma_{\rm in}(n_{\rm gwb}-4)}}{\langle e^{-\Gamma_{\rm in}(n_{\rm gwb}-4)} \rangle}-1\,.
\end{equation} 
\subsection{Graviton geodesic equation}
The inhomogeneities due to the propagation of gravitons through the universe are contained in the ratio of the comoving momenta at emission and at observation, which can be derived from the time component of the geodesic equation for gravitons
\begin{equation}
    \frac{d p^{0}}{d\eta} = -\Gamma^{0}_{\nu\lambda} \frac{p^{\nu}p^{\lambda}}{p^{0}}\,.
\end{equation} 
Gravitons propagate along null geodesics 
\begin{equation}
    p^2 = g_{\mu\nu}p^\mu p^\nu = 0 \, ,
\end{equation}
then we can write
\begin{equation}
    -a^2e^{2\Psi}(p^0)^2 + p^2  = 0 \,,
\end{equation}
where we defined 
\begin{equation}
    p^2 \equiv g_{ij}p^ip^j\,.
\end{equation}
This gives
\begin{equation}
\begin{split}
    p^i = \frac{q}{a^2}\hat{n}^i e^{\Phi}\,,\quad\quad\quad\quad  p^0 = \frac{q}{a^2}e^{-\Psi} \,.
\end{split}
\end{equation}
The left-hand side of the geodesic equation is equal to 
\begin{equation}
    \frac{dp^0}{d\eta} = -2\mathcal{H}\frac{q}{a^2}e^{-\Psi}  - \Psi^\prime \frac{q}{a^2}e^{-\Psi} + \frac{dq}{d\eta} \frac{1}{a^2}e^{-\Psi}\,. 
\end{equation}
Consequently, the geodesic equation reads
\begin{equation}
     \frac{1}{q}\frac{dq}{d\eta} =\Phi^\prime -n^i\partial_i\Psi e^{\Psi+\Phi}
\end{equation}
which can be rewritten as 
\begin{equation}
   \frac{1}{q} \frac{dq}{d\eta} = \Phi^\prime + \Psi^\prime - \frac{d\Psi}{d\eta} \, .
\end{equation}
After integration, we get the relation between the comoving momenta at the emission and observation times
\begin{equation}
    q_{\rm 0} = q_{\rm in}e^{\Psi(\eta_{\rm in},\Vec{x}) - \Psi(\eta_{0},\Vec{x}) + \int^{\eta_{0}}_{\eta_{\rm in}} d\eta \left(\Phi^\prime(\eta,\Vec{x}) + \Psi^\prime(\eta,\Vec{x})\right)} \, ,
\end{equation}
The first and the third term represent the SW and ISW effects respectively, while the second term contributes only to the monopole term and therefore can be disregarded. Substituting this ratio and the initial conditions in eq. \eqref{ratio}
we obtain the non-perturbative generalization of the CGWB anisotropy on large scales
\begin{equation}
    \Gamma_{0}  =  \frac{-2\Psi(\eta_{\rm in},\Vec{x})+ 4\Phi(\eta_{\rm in},\Vec{x})}{4-n_{\rm gwb}(q)} 
     + \Psi(\eta_{\rm in},\Vec{x})  + \int^{\eta_{0}}_{\eta_{\rm in}} d\eta \left(\Phi^\prime(\eta,\Vec{x}) + \Psi^\prime(\eta,\Vec{x})\right)\, \, .
\end{equation} 
Due to the exponentiation of gravitational potentials in the definition of the metric this result is fully non-linear. From now on, we will restrict our analysis to the initial conditions and the SW effect only, leaving the analysis of the ISW contribution to a follow-up study.

\section{The CGWB bispectrum } 
The non-perturbative initial conditions and SW effect can be expressed in terms of the comoving curvature perturbation $\zeta$. This will allow us to relate the anisotropy of the CGWB today directly to the initial conditions determined at the end of inflation. Assuming that all fluids have a definite equation of state, from the continuity equation we derive the following quantity~\footnote{{ We assume that adiabaticity holds only for standard radiation and matter perturbations and the GW contribution to the total energy density is neglibigle.}}~\cite{Kolb:2004jg,Lyth:2004gb}
\begin{equation}
    \mathcal{F} = \ln{\Tilde{a}}  + \int^{\Tilde{\rho}}_{\rho}\frac{d\Tilde{\rho}'}{\Tilde{\rho}' + \Tilde{P}'}\,,
\end{equation}
where the tilde refers to the redefined scale factor in eq. \eqref{eq:red_metric}.
This quantity is conserved at any order in perturbation theory and $\delta F = \zeta$ represents a non-linear generalization of the comoving curvature perturbation $\zeta$ on superhorizon scales. Using eqs. \eqref{redef}, we can write  
\begin{equation}
\label{zeta}
    \zeta = \delta F = - \Phi  + \frac{1}{4}\ln{\frac{\Tilde{\rho}}{\rho}} =  - \Phi  - \frac{1}{2}\Psi \,.
\end{equation}
where we have used the equation of state, $w = 1/3$, since the future GW detectors, such as LISA~\cite{LISA:2022kgy}, Einstein Telescope~\cite{Branchesi:2023mws}, Cosmic Explorer~\cite{Reitze:2019iox}, BBO~\cite{Corbin:2005ny}, DECIGO~\cite{Kawamura:2006up}, are expected to probe the frequency range corresponding to modes that reenter the horizon during the radiation-dominated era. The total energy density defined in the new metric can be inferred from the (0,0) component of Einstein equations, $\Tilde{\rho} =  e^{-2\Psi} \, \rho$. 

The scalar perturbations $\Phi$ and $\Psi$ are related by the transverse-traceless part of the Einstein equation in the Poisson gauge, which at higher order in perturbations includes the non-local terms
\begin{equation}
\begin{split}
    \nabla^4\left(\Phi-\Psi\right) &= - \frac{3}{2}\partial_i\partial^k(\partial^i\Phi\partial_k\Phi) + \frac{1}{2}\nabla^2(\partial^i\Phi\partial_i\Phi) - 3\partial_i\partial^k(\partial^i\Psi\partial_k\Psi) + \nabla^2(\partial^i\Psi\partial_i\Psi)\\
    &  +3\partial_i\partial^k(\partial^i\Psi\partial_k\Phi) -\nabla^2(\partial^i\Psi\partial_k\Phi)  \, .
\end{split}
\end{equation}
We can take into account the non-local terms by introducing a suitable kernel $\mathcal{K}[\Psi, \Phi]$. Then, it is possible to write 
\begin{equation}
\label{psi}
    \Psi = \Phi + \mathcal{K}[\Psi, \Phi] \, .
\end{equation}
Using eqs. \eqref{zeta} and \eqref{psi}, the perturbation of the graviton distribution function reads
\begin{equation}
 \Gamma_0= -\frac{2}{3} \frac{6-n_{\rm gwb}}{4-n_{\rm gwb}}\zeta -\frac{2}{3}\frac{n_{\rm gwb}}{4-n_{\rm gwb}}\mathcal{K} \,.
\end{equation} 
In the case of a scale-invariant power spectrum, $n_{\rm gwb}=0$, the dependence on the kernel vanishes, and we are left with
\begin{equation}
 \Gamma_0= -\zeta \,.
\end{equation}  
 While in the CMB case, the non-local contribution to large-scale anisotropy disappears only in the squeezed limit, for the inflationary CGWB instead this is valid for every configuration. The full non-linearity is contained in the curvature perturbation. Since we are interested in computing the 3-point correlation function of the GW energy density perturbation, we can write
\begin{equation}
\label{gw delta}
    \delta_{\rm GW} = A \, e^{-\zeta} -1 \, ,
\end{equation}
where $A = 1/ \langle e^{\Gamma_0}\rangle$, so that the the GW energy density perturbation is equal to zero if we take an ensamble average.

To compute the n-point correlation function, we can use the path-integral approach~\cite{Ramond:1981pw, Zinn-Justin:1989rgp}. Introducing the generating functional of the correlation functions defined as
\begin{equation}
    Z[J] = \int D[\zeta] P[\zeta] e^{i\int d\Vec{x} J(\Vec{x})\left(A \, e^{-\zeta(\Vec{x})}-1 \right)}\,,
\end{equation} 
where $J(\Vec{x})$ is an external source perturbing the underlying statistics, $P[\zeta]$ is a probability distribution, and $D[\zeta]$ is an appropriate measure such that $\int D[\zeta] P[\zeta] = 1$. By taking the functional derivatives of $Z[J]$, we can get the n-point correlation function
\begin{equation}
   \langle (A \, e^{-\zeta_1} -1)...(A \, e^{-\zeta_n} -1)\rangle = i^{-n} \frac{\delta^n Z[J]}{\delta J(x_1)...\delta J(x_n)} = Z^{(n)}(x_1,...,x_n)\,,
\end{equation}
where $\zeta(\Vec{x}_i)= \zeta_i$. Defining a new functional $W \equiv \ln{Z[J]}$ allows us to obtain the connected correlation function
\begin{equation}
      \langle (A \, e^{-\zeta_1} -1)...(A \, e^{-\zeta_n} -1)\rangle_{\rm conn.} = i^{-n} \frac{\delta^n W[J]}{\delta J(x_1)...\delta J(x_n)} = W^{(n)}(x_1,...,x_n)\,.
\end{equation} 
In the case of standard single-field models of inflation, the amount of primordial non-Gaussianity is negligible and the only contribution to non-Gaussianity comes from the post-inflationary non-linear evolution of cosmological perturbations. Therefore, we assume $\zeta$ to be a Gaussian distributed quantity.  In this case, the GW energy density perturbation $\delta_{\rm GW}$ will be lognormally distributed~\cite{ 10.1093/mnras/228.2.407,Xavier:2016elr}, which gives rise to such effect as intermittency, consisting in the emergence of
of isolated high density spots, separated by large underdense areas~\cite{Coles:2001fw,YaBZel'dovich_1987,Schandarin:1989sr}, which would make the cosmological signal distinguishable from the astrophysical one. 

Using the property of the n-point correlation function for the exponent of the Gaussian random field~\cite{Bartolo:2005fp}, we can obtain the 2-point connected correlation function for the GW energy density perturbation
\begin{equation}
    W^{(2)}(x_1, x_2) = \langle (A \, e^{-\zeta_1} -1)(A \, e^{-\zeta_2} -1)\rangle_{\rm conn}= e^{\langle\zeta_1\zeta_2\rangle}-1 \,.
\end{equation}
Consequently, the connected contribution to the exact 3-point correlation function is 
\begin{equation}
\begin{split}
     W^{(3)}(x_1, x_2, x_3) &= \langle (A \, e^{-\zeta_1} -1) (A \, e^{-\zeta_2} -1)(A \, e^{-\zeta_3} -1)\rangle_{\rm conn} = \\ &e^{\langle\zeta_1\zeta_2\rangle +\langle\zeta_2\zeta_3\rangle+ \langle\zeta_1\zeta_3\rangle}-e^{\langle\zeta_1\zeta_2\rangle} - e^{\langle\zeta_2\zeta_3\rangle} -e^{\langle\zeta_1\zeta_3\rangle} + 2 \, .
     \end{split}
\end{equation}
which can be rewritten as 
\begin{equation}
\begin{split}  
     W^{(3)}(x_1, x_2, x_3) &= W^{(2)}(x_1, x_2)W^{(2)}(x_2, x_3)W^{(2)}(x_1, x_3)  +  W^{(2)}(x_1, x_2)W^{(2)}(x_2, x_3) \\&+ W^{(2)}(x_2, x_3)W^{(2)} (x_1, x_3) + W^{(2)}(x_1, x_2)W^{(2)}(x_1, x_3) \,.
\end{split}
\end{equation}
To take into account a possible primordial non-Gaussian contribution, the curvature perturbations in the case of local non-Gaussianity can be expanded as
\begin{equation}
    \zeta = \zeta_{\rm G} + \frac{3}{5}f_{\rm NL}\zeta_{G}^2\, ,
\end{equation}
where $ \zeta_{\rm G}$ is a linear Gaussian part and $f_{\rm NL}$ parametrizes the amount of primordial non-Gaussianity. Expanding eq. \eqref{gw delta} up to the second order in $\zeta$, we can obtain the effective non-linearity parameter $f_{\rm NL}^{\rm GW}$ for the bispectrum of the GW energy density~\cite{Gangui:1993tt, Komatsu:2000vy}
\begin{equation} 
    f_{\rm NL}^{\rm GW} = f_{\rm NL} - \frac{5}{6} \,.
\end{equation}
Hence, the main contribution to the non-Gaussianity in standard single-field inflation comes from the post-inflationary evolution of gravitational potentials.

\section{Conclusion}
Since General Relativity is fundamentally non-linear, higher-order corrections leave their imprint on cosmological perturbations. In this work, we have presented a non-perturbative approach for the computation of the CGWB anisotropy on large scales. Anisotropies in the GW energy density arise at the moment of emission and further during their propagation. The initial anisotropy is strongly dependent on the production mechanism~\cite{Schulze:2023ich, ValbusaDallArmi:2024hwm} (see also \cite{Iovino:2024sgs} for a non-perturbative calculation applied in the Primordial Black Hole context). We considered the CGWB produced by the quantum fluctuation of the metric during inflation and derived the non-perturbative generalization of the initial anisotropy of the CGWB by rescaling the large-scale scalar and small-scale tensor perturbations extending the result of~\cite{ValbusaDallArmi:2023nqn,ValbusaDallArmi:2024hwm} for the initial conditions of the CGWB. We obtained the anisotropic contribution due to the propagation of GWs through the large-scale scalar perturbations solving the geodesic equation for the gravitons.

An important aspect of the non-linear effects is the deviation of the primordial perturbations from Gaussian statistics. On large scales, the initial conditions and the SW effect  provides the leading contribution to the CGWB anisotropy, however the late ISW eﬀect is also significant. Conversely, effects such as gravitational lensing and Shapiro time-delay become important on smaller scales. The non-linear CGWB anisotropies can affect higher-order correlation functions. The bispectrum of the CGWB is a powerful tool to measure non-Gaussianity on cosmological scales. Besides the primordial contribution generated during or right after inflation, the non-Gaussianity of the CGWB is caused by the non-linear evolution of gravitational potentials. Using the expression for the CGWB anisotropy in terms of the comoving curvature perturbation, we have computed non-perturbatively the 3-point correlation function of the GW energy density. An important result of this work is that in the case of CGWB generated by quantum tensor fluctuations during inflation with the scale-invariant power spectrum and negligible primordial non-Gaussianity, the contribution from the kernel $\mathcal{K}$, which contains further non-Gaussian contributions, vanishes from the expression of the GW energy density perturbation. This will make the SW effect of the CGWB energy density completely insensitive of the possible presence of gravitational slip (i.e., $\Phi\neq \Psi$) even beyond the squeezed limit (as opposed to the adiabatic initial conditions) as arising from e.g. the anisotropic stress
produced by collisionless relativistic particles like neutrinos or by modifications of gravity.
The effects coming from the presence of anisotropic stress will instead remain in the ISW terms.
Finally, we have accounted also for (local) primordial non-Gaussianity, providing the perturbative expression for the effective non-linearity parameter  $f_{\rm NL}$ characterizing the quadratic non-linearity in the CGWB anisotropies.

\section*{Acknowledgments}
N.B. and S.M. acknowledge support from the COSMOS network
(www.cosmosnet.it) through the
ASI (Italian Space Agency) Grants 2016-24-H.0, 2016-24-H.1-2018 and
2020-9-HH.0.

\bibliographystyle{ieeetr}
\bibliography{bibliography.bib}

\begin{thebibliography}{10}

\bibitem{NANOGrav:2023gor}
G.~Agazie {\em et~al.}, ``{The NANOGrav 15 yr Data Set: Evidence for a Gravitational-wave Background},'' {\em Astrophys. J. Lett.}, vol.~951, no.~1, p.~L8, 2023.

\bibitem{EPTA:2023fyk}
J.~Antoniadis {\em et~al.}, ``{The second data release from the European Pulsar Timing Array - III. Search for gravitational wave signals},'' {\em Astron. Astrophys.}, vol.~678, p.~A50, 2023.

\bibitem{Reardon:2023gzh}
D.~J. Reardon {\em et~al.}, ``{Search for an Isotropic Gravitational-wave Background with the Parkes Pulsar Timing Array},'' {\em Astrophys. J. Lett.}, vol.~951, no.~1, p.~L6, 2023.

\bibitem{NANOGrav:2023hvm}
A.~Afzal {\em et~al.}, ``{The NANOGrav 15 yr Data Set: Search for Signals from New Physics},'' {\em Astrophys. J. Lett.}, vol.~951, no.~1, p.~L11, 2023.

\bibitem{NANOGrav:2023hfp}
G.~Agazie {\em et~al.}, ``{The NANOGrav 15 yr Data Set: Constraints on Supermassive Black Hole Binaries from the Gravitational-wave Background},'' {\em Astrophys. J. Lett.}, vol.~952, no.~2, p.~L37, 2023.

\bibitem{EPTA:2023xxk}
J.~Antoniadis {\em et~al.}, ``{The second data release from the European Pulsar Timing Array - IV. Implications for massive black holes, dark matter, and the early Universe},'' {\em Astron. Astrophys.}, vol.~685, p.~A94, 2024.

\bibitem{Franciolini:2023pbf}
G.~Franciolini, A.~Iovino, Junior., V.~Vaskonen, and H.~Veermae, ``{Recent Gravitational Wave Observation by Pulsar Timing Arrays and Primordial Black Holes: The Importance of Non-Gaussianities},'' {\em Phys. Rev. Lett.}, vol.~131, no.~20, p.~201401, 2023.

\bibitem{Franciolini:2023wjm}
G.~Franciolini, D.~Racco, and F.~Rompineve, ``{Footprints of the QCD Crossover on Cosmological Gravitational Waves at Pulsar Timing Arrays},'' {\em Phys. Rev. Lett.}, vol.~132, no.~8, p.~081001, 2024.

\bibitem{Ellis:2023tsl}
J.~Ellis, M.~Lewicki, C.~Lin, and V.~Vaskonen, ``{Cosmic superstrings revisited in light of NANOGrav 15-year data},'' {\em Phys. Rev. D}, vol.~108, no.~10, p.~103511, 2023.

\bibitem{Vagnozzi:2023lwo}
S.~Vagnozzi, ``{Inflationary interpretation of the stochastic gravitational wave background signal detected by pulsar timing array experiments},'' {\em JHEAp}, vol.~39, pp.~81--98, 2023.

\bibitem{Figueroa:2023zhu}
D.~G. Figueroa, M.~Pieroni, A.~Ricciardone, and P.~Simakachorn, ``{Cosmological Background Interpretation of Pulsar Timing Array Data},'' {\em Phys. Rev. Lett.}, vol.~132, no.~17, p.~171002, 2024.

\bibitem{Cornish:2001hg}
N.~J. Cornish, ``{Mapping the gravitational wave background},'' {\em Class. Quant. Grav.}, vol.~18, pp.~4277--4292, 2001.

\bibitem{Alba:2015cms}
V.~Alba and J.~Maldacena, ``{Primordial gravity wave background anisotropies},'' {\em JHEP}, vol.~03, p.~115, 2016.

\bibitem{Contaldi:2016koz}
C.~R. Contaldi, ``{Anisotropies of Gravitational Wave Backgrounds: A Line Of Sight Approach},'' {\em Phys. Lett. B}, vol.~771, pp.~9--12, 2017.

\bibitem{LISACosmologyWorkingGroup:2022kbp}
N.~Bartolo {\em et~al.}, ``{Probing anisotropies of the Stochastic Gravitational Wave Background with LISA},'' {\em JCAP}, vol.~11, p.~009, 2022.

\bibitem{Bartolo:2019oiq}
N.~Bartolo, D.~Bertacca, S.~Matarrese, M.~Peloso, A.~Ricciardone, A.~Riotto, and G.~Tasinato, ``{Anisotropies and non-Gaussianity of the Cosmological Gravitational Wave Background},'' {\em Phys. Rev. D}, vol.~100, no.~12, p.~121501, 2019.

\bibitem{ValbusaDallArmi:2020ifo}
L.~Valbusa~Dall'Armi, A.~Ricciardone, N.~Bartolo, D.~Bertacca, and S.~Matarrese, ``{Imprint of relativistic particles on the anisotropies of the stochastic gravitational-wave background},'' {\em Phys. Rev. D}, vol.~103, no.~2, p.~023522, 2021.

\bibitem{Schulze:2023ich}
F.~Schulze, L.~Valbusa~Dall'Armi, J.~Lesgourgues, A.~Ricciardone, N.~Bartolo, D.~Bertacca, C.~Fidler, and S.~Matarrese, ``{GW\_CLASS: Cosmological Gravitational Wave Background in the cosmic linear anisotropy solving system},'' {\em JCAP}, vol.~10, p.~025, 2023.

\bibitem{Bartolo:2019zvb}
N.~Bartolo, D.~Bertacca, V.~De~Luca, G.~Franciolini, S.~Matarrese, M.~Peloso, A.~Ricciardone, A.~Riotto, and G.~Tasinato, ``{Gravitational wave anisotropies from primordial black holes},'' {\em JCAP}, vol.~02, p.~028, 2020.

\bibitem{Ricciardone:2021kel}
A.~Ricciardone, L.~V. Dall'Armi, N.~Bartolo, D.~Bertacca, M.~Liguori, and S.~Matarrese, ``{Cross-Correlating Astrophysical and Cosmological Gravitational Wave Backgrounds with the Cosmic Microwave Background},'' {\em Phys. Rev. Lett.}, vol.~127, no.~27, p.~271301, 2021.

\bibitem{Bartolo:2019yeu}
N.~Bartolo, D.~Bertacca, S.~Matarrese, M.~Peloso, A.~Ricciardone, A.~Riotto, and G.~Tasinato, ``{Characterizing the cosmological gravitational wave background: Anisotropies and non-Gaussianity},'' {\em Phys. Rev. D}, vol.~102, no.~2, p.~023527, 2020.

\bibitem{Perna:2023dgg}
G.~Perna, A.~Ricciardone, D.~Bertacca, and S.~Matarrese, ``{Non-Gaussianity from the cross-correlation of the astrophysical Gravitational Wave Background and the Cosmic Microwave Background},'' {\em JCAP}, vol.~10, p.~014, 2023.

\bibitem{Pyne:1995bs}
T.~Pyne and S.~M. Carroll, ``{Higher order gravitational perturbations of the cosmic microwave background},'' {\em Phys. Rev. D}, vol.~53, pp.~2920--2929, 1996.

\bibitem{Mollerach:1997up}
S.~Mollerach and S.~Matarrese, ``{Cosmic microwave background anisotropies from second order gravitational perturbations},'' {\em Phys. Rev. D}, vol.~56, pp.~4494--4502, 1997.

\bibitem{Bartolo:2011wb}
N.~Bartolo, S.~Matarrese, and A.~Riotto, ``{Non-Gaussianity in the Cosmic Microwave Background Anisotropies at Recombination in the Squeezed limit},'' {\em JCAP}, vol.~02, p.~017, 2012.

\bibitem{Boubekeur:2009uk}
L.~Boubekeur, P.~Creminelli, G.~D'Amico, J.~Norena, and F.~Vernizzi, ``{Sachs-Wolfe at second order: the CMB bispectrum on large angular scales},'' {\em JCAP}, vol.~08, p.~029, 2009.

\bibitem{Bartolo:2005fp}
N.~Bartolo, S.~Matarrese, and A.~Riotto, ``{Non-Gaussianity of Large-Scale Cosmic Microwave Background Anisotropies beyond Perturbation Theory},'' {\em JCAP}, vol.~08, p.~010, 2005.

\bibitem{Roldan:2017wvm}
O.~Roldan, ``{CMB anisotropies at all orders: the non-linear Sachs-Wolfe formula},'' {\em JCAP}, vol.~08, p.~034, 2017.

\bibitem{Cai:2024dya}
R.-G. Cai, S.-J. Wang, Z.-Y. Yuwen, and X.-X. Zeng, ``{Anisotropies of cosmological gravitational wave backgrounds in non-flat spacetime},'' 10 2024.

\bibitem{LISACosmologyWorkingGroup:2024hsc}
M.~Braglia {\em et~al.}, ``{Gravitational waves from inflation in LISA: reconstruction pipeline and physics interpretation},'' {\em JCAP}, vol.~11, p.~032, 2024.

\bibitem{ValbusaDallArmi:2023nqn}
L.~Valbusa~Dall'Armi, A.~Mierna, S.~Matarrese, and A.~Ricciardone, ``{Adiabatic or Non-Adiabatic? Unraveling the Nature of Initial Conditions in the Cosmological Gravitational Wave Background},'' 7 2023.

\bibitem{ValbusaDallArmi:2024hwm}
L.~Valbusa~Dall'Armi, A.~Mierna, S.~Matarrese, and A.~Ricciardone, ``{Inflationary initial conditions for the cosmological gravitational wave background},'' {\em JCAP}, vol.~07, p.~043, 2024.

\bibitem{Galloni:2022mok}
G.~Galloni, N.~Bartolo, S.~Matarrese, M.~Migliaccio, A.~Ricciardone, and N.~Vittorio, ``{Updated constraints on amplitude and tilt of the tensor primordial spectrum},'' {\em JCAP}, vol.~04, p.~062, 2023.

\bibitem{PhysRevD.42.3936}
D.~S. Salopek and J.~R. Bond, ``Nonlinear evolution of long-wavelength metric fluctuations in inflationary models,'' {\em Phys. Rev. D}, vol.~42, pp.~3936--3962, Dec 1990.

\bibitem{Isaacson:1967sln}
R.~A. Isaacson, {\em {Gravitational Radiation in the Limit of High Frequency}}.
\newblock PhD thesis, Maryland U., 1967.

\bibitem{Kolb:2004jg}
E.~W. Kolb, S.~Matarrese, A.~Notari, and A.~Riotto, ``{Cosmological influence of super-Hubble perturbations},'' {\em Mod. Phys. Lett. A}, vol.~20, pp.~2705--2710, 2005.

\bibitem{Lyth:2004gb}
D.~H. Lyth, K.~A. Malik, and M.~Sasaki, ``{A General proof of the conservation of the curvature perturbation},'' {\em JCAP}, vol.~05, p.~004, 2005.

\bibitem{LISA:2022kgy}
K.~G. Arun {\em et~al.}, ``{New horizons for fundamental physics with LISA},'' {\em Living Rev. Rel.}, vol.~25, no.~1, p.~4, 2022.

\bibitem{Branchesi:2023mws}
M.~Branchesi {\em et~al.}, ``{Science with the Einstein Telescope: a comparison of different designs},'' {\em JCAP}, vol.~07, p.~068, 2023.

\bibitem{Reitze:2019iox}
D.~Reitze {\em et~al.}, ``{Cosmic Explorer: The U.S. Contribution to Gravitational-Wave Astronomy beyond LIGO},'' {\em Bull. Am. Astron. Soc.}, vol.~51, no.~7, p.~035, 2019.

\bibitem{Corbin:2005ny}
V.~Corbin and N.~J. Cornish, ``{Detecting the cosmic gravitational wave background with the big bang observer},'' {\em Class. Quant. Grav.}, vol.~23, pp.~2435--2446, 2006.

\bibitem{Kawamura:2006up}
S.~Kawamura {\em et~al.}, ``{The Japanese space gravitational wave antenna DECIGO},'' {\em Class. Quant. Grav.}, vol.~23, pp.~S125--S132, 2006.

\bibitem{Ramond:1981pw}
P.~Ramond, {\em {Field theory: a modern primer}}, vol.~51.
\newblock 1981.

\bibitem{Zinn-Justin:1989rgp}
J.~Zinn-Justin, {\em {Quantum field theory and critical phenomena}}, vol.~77 of {\em International Series of Monographs on Physics}.
\newblock Oxford University Press, 4 2021.

\bibitem{10.1093/mnras/228.2.407}
P.~Coles and J.~D. Barrow, ``Non-gaussian statistics and the microwave background radiation,'' {\em Monthly Notices of the Royal Astronomical Society}, vol.~228, pp.~407--426, 09 1987.

\bibitem{Xavier:2016elr}
H.~S. Xavier, F.~B. Abdalla, and B.~Joachimi, ``{Improving lognormal models for cosmological fields},'' {\em Mon. Not. Roy. Astron. Soc.}, vol.~459, no.~4, pp.~3693--3710, 2016.

\bibitem{Coles:2001fw}
P.~Coles, ``{The Origin of spatial intermittency in the galaxy distribution},'' {\em Mon. Not. Roy. Astron. Soc.}, vol.~330, p.~421, 2002.

\bibitem{YaBZel'dovich_1987}
Y.~B. Zel'dovich, S.~A. Molchanov, A.~A. Ruzmaĭkin, and D.~D. Sokolov, ``Intermittency in random media,'' {\em Soviet Physics Uspekhi}, vol.~30, p.~353, may 1987.

\bibitem{Schandarin:1989sr}
S.~F. Schandarin and Y.~B. Zeldovich, ``{The Large scale structure of the universe: Turbulence, intermittency, structures in a selfgravitating medium},'' {\em Rev. Mod. Phys.}, vol.~61, pp.~185--220, 1989.

\bibitem{Gangui:1993tt}
A.~Gangui, F.~Lucchin, S.~Matarrese, and S.~Mollerach, ``{The Three point correlation function of the cosmic microwave background in inflationary models},'' {\em Astrophys. J.}, vol.~430, pp.~447--457, 1994.

\bibitem{Komatsu:2000vy}
E.~Komatsu and D.~N. Spergel, ``{The Cosmic Microwave Background bispectrum as a test of the physics of inflation and probe of the astrophysics of the low-redshift Universe},'' in {\em {9th Marcel Grossmann Meeting (MG 9)}}, pp.~2009--2012, 12 2000.

\bibitem{Iovino:2024sgs}
A.~J. Iovino, S.~Matarrese, G.~Perna, A.~Ricciardone, and A.~Riotto, ``{How Well Do We Know the Scalar-Induced Gravitational Waves?},'' 12 2024.

\end{thebibliography}

\end{document}